# Gap opening and orbital modification of superconducting FeSe above the structural distortion

Y.-C. Wen,[1] K.-J. Wang,[1] H.-H. Chang,[2] J.-Y. Luo,[1] C.-C. Shen,[3] H.-L. Liu,[3] C.-K. Sun,[1,4] M.-J. Wang,[2] and M.-K. Wu[1,*]

[1]*Institute of Physics, Academia Sinica, Taipei 115, Taiwan*

[2]*Institute of Astronomy and Astrophysics, Academia Sinica, Taipei 10617, Taiwan*

[3]*Department of Physics, National Taiwan Normal University, Taipei 116, Taiwan*

[4]*Molecular Imaging Center and Department of Electrical Engineering and Graduate Institute of Photonics and Optoelectronics, National Taiwan University, Taipei 10617, Taiwan*

We utilize steady-state and transient optical spectroscopies to examine the responses of nonthermal quasiparticles with respect to orbital modifications in normal-state iron-chalcogenide superconductors. The dynamics shows the emergence of gap-like quasiparticles (associated to a ~36 meV energy gap) with a coincident transfer of the optical spectral weight in the visible range, at temperatures *above* the structural distortion. Our observations suggest that opening of the high-temperature gap and the lattice symmetry breaking are possibly driven by short-range orbital and/or charge orders, implicating a close correlation between electronic nematicity and precursor order in iron-based superconductors.

Discovery of iron-based superconductors is of fundamental importance as it may provide valuable insight to better understand high-transition-temperature (high-$T_c$) superconductivity [1-3]. In this new class of materials, electron pairing was suggested to be mediated by spin fluctuations [3,4] and likely entangled in the electronic nematic order (a rotational symmetry-breaking phase) [5-8]. Both nematicity and magnetic fluctuations are generally associated with the structural distortion at high temperatures. The interplay among different degrees of freedom is thus the core of diverse questions involving Fe-based superconductors. Key ongoing debates are how the nematicity develops and whether it induces precursory electron condensations [9]. Very recently, angle-resolved photoemission spectroscopic studies [5] indicated that breaking of the electronic symmetry is induced by unbalanced populations in Fe $3d_{xz}$ and $3d_{yz}$ orbitals, driven by the orbital splitting near the structural transition temperature ($T_s$). This effect causes changes in transport properties through Fermi surface reconstruction [9], and redistributions of the optical spectral weight with photon energies up to the visible range [7,8]. Even with these efforts, no generally accepted picture has emerged due to limited observations on the orbital modification at temperatures across $T_s$ [5]. It thus remains intricate to address the role which nematicity plays in relation to the lattice symmetry breaking, as well as high-$T_c$ superconductivity.

Time-resolved optical spectroscopy [10-12], an energy-resolved and bulk probe into electron and phonon dynamics, provides a delicate insight into changes of the band structure near the Fermi level $E_F$ and underlying interatomic interactions. Indeed, these studies on cuprate [10] and iron-pnictide (1111- and 122-type) compounds [11,12] have revealed the existence of an energy gap above $T_c$ and argued its responsible role in the precursor order. However, is the opening of this high-temperature gap driven by the anisotropic charge ordering? Is the gap ubiquitous in all types of iron-based superconductors?

In this Letter, we report investigations on quasiparticle and acoustic phonon dynamics with respect to the orbital modification in iron chalcogenides $FeSe_{1-x}$, using steady-state and transient optical spectroscopies. Relaxation of quasiparticles across a ~36 meV energy gap is observed at $T < 130$ K, accompanying with a coincident transfer of the optical spectral weight in the visible range and alterations in the transport properties. The *in situ* and strain-free observations provide compelling evidence that the modification of the electronic structure is *prior to* the lattice transformation. Our results suggest that the high-temperature gap and the lattice symmetry breaking are driven by short-range orbital and/or charge orders, implicating a close correlation between nematicity and the precursor phase.

Tetragonal $FeSe_{1-x}$ consists of two-dimensional layers of edge-sharing Fe-Se tetrahedra. The lack of interlayer spacing atom makes the exchange interactions between the primary elements manifest, presumably clarifying orbital-associated observations. Together with the absence of long-range magnetic order [4], FeSe provides the simplest platform to examine the structural and orbital modifications. The sample under study was a 460 nm-thick (101) $FeSe_{1-x}$ thin film grown on a (001) MgO substrate using pulsed laser deposition. Detailed sample characterization confirmed the single crystallinity of the nearly stoichiometric $FeSe_{1-x}$ without noticeable substrate-induced strain effect [13]. *T*-dependent resistivity indicated an onset $T_c$ of 10 K and a slope change near ~90 K, as an indication of the structural distortion [2]. The lattice structure of the $FeSe_{1-x}$ has been studied and reported previously [13,14]. In parts of the presented experiments, a 450 nm-thick (001) FeSe film was also used as a reference for inspecting anisotropic properties.

The major contribution to the optical response of FeSe in the visible spectral range comes from intraband Fe *d* and interband Fe *d* to hybridized Fe-*d*/Se-*p* transitions. Evolutions of these orbitals can thus be revealed by steady-state optical spectroscopies [7,8]. Fig. 1 shows the real part of the optical conductivity $\sigma_1(\omega)$, taken from room-temperature ellipsometric measurements. The comparison between the measured $\sigma_1(\omega)$ and the density functional calculation [15] allows us to attribute the distinct peak at ~1.8 eV of the conductivity spectrum to the transition from Fe $d$ to Fe-$d$/Se-$p$ hybridized orbitals. To look into the $T$-dependent orbital change, we used near infrared with a photon energy of 1.55 eV (an arrow in Fig. 1) to resonantly probe this interband transition. Inset of Fig. 1 shows the corresponding $T$-dependent optical reflectivity, where a clear reduction can be found at $T <$ 130 ± 10 K (a marker $T_e$ is set to be 140 K). This manifest feature indicates a redistribution of the optical spectral weight, serving as the evidence for the orbital modification at low temperatures.

Transient pump-probe experiments were performed to explore possible energy gaps near $E_F$ in the proximity to $T_e$. The light source was a mode-locked Ti:sapphire laser with 76 MHz repetition rate and 170 fs pulse width. It output 1.55 eV photons for the optical probe, corresponding to the optimized sensitivity to the orbital change, while the optical pump was frequency-doubled. We kept the pump fluence 5.3 μJ/cm$^2$ for examining the dynamics in the weak perturbative regime [10], wherein all contributions to the reflectivity changes ΔR/R can be assumed linear in the photoexcited carrier density. The resultant small ΔR/R (10$^{-6}$) was discerned using synchronous detection with a modulation frequency of 400 kHz. The sample was attached on a heat exchanger with variable temperatures of 20 – 297 K.

After the photoexcitation, the pump-induced nonthermal carriers soon reach their internal equilibrium through efficient carrier-carrier scatterings, followed by external thermalization with the phonon bath. Sub-nanosecond recovery of the system is then governed by the spin-lattice relaxation and the heat diffusion into the substrate. To examine changes of the electronic structure near $E_F$, we first inspect the carrier thermalization and recombination processes which are revealed by the picosecond response of ΔR/R [Fig. 2(a)]. It is found that the high-$T$ traces can be well described by a single relaxation (straight line in logarithm plot), whereas an additional sub-ps relaxation is observed at low $T$. This indicates that a two-component fit to the data is necessary for an accurate description. We thus model the picosecond response by $\Delta R/R = A_{fast} \exp(-t/\tau_{fast}) + A_{slow} \exp(-t/\tau_{slow}) + A_{step}$, where additional $A_{step}$ represents the sub-ns recovery.

Fig. 2(b) and (c) show the $T$-dependent fitting variables. From the inset of Fig. 2(b), the sub-ps relaxation (*fast* one) is found to emerge below 130 K, and its magnitude $A_{fast}$ gradually increases with decreasing $T$ until ~70 K. The feature allows us to attribute this signal to the gap-like quasiparticles, which can be phenomenologically explained in the framework of Rothwarf-Taylor (RT) theory [16] - In the presence of a gap near $E_F$, hot carriers will accumulate in the quasiparticle state above the gap and wait for recombination via scatterings with high-energy phonons.

Additionally, the picosecond relaxation (*slow* one) is contributed from carrier-phonon (*c-p*) thermalization that is ubiquitous at all $T$ [11,12]. Intriguingly, we find that the measured *c-p* thermalization rate strongly depends on the film orientation, indicating anisotropy of the probe transition matrix element. For a probe polarization parallel to the *ab* plane (see supplementary information [17]), the *c-p* thermalization rate is 2.7 ps$^{-1}$ in a reasonable agreement with the theoretical prediction (3.5 ps$^{-1}$) [17]. This process is found to be extremely slow as probing FeSe with a notable polarization component along *c*-axis, e.g., normally-incident probe of (101) film [$\tau_{slow}^{-1}$ ~0.2 ps$^{-1}$ in Fig. 2(b)]. This striking contrast reflects a dramatic change in the contribution of bands with non-vanishing interplane wavevectors to $\Delta R/R$ and considerable mass of the carriers therein [18]. We speculate that the slow *c-p* thermalization observed in the (101) FeSe is mainly arising from carriers in the nearly-dispersionless hole-like band around $\Gamma - Z$ line in adjacent to $E_F$ [15]. The disclosed anisotropic *c-p* coupling could lead to the perception of distinctive interplane optical conductivity and resistivity [19,20].

Theoretical analyses of the two relaxations are carried out to acquire better understanding of the gap. Details of the adopted theories and the data fitting can be found in Ref. 17 and Fig. 2(b), and the conclusions are given as follows. First, the two-temperature model (TTM) [21], providing the commonly accepted descriptions of the *c-p* coupling in metals, predicts an overestimated temperature dependency ($\propto T^2 \sim T^3$) of $\tau_{slow}^{-1}$ at low *T*. This discrepancy can be alleviated by considering a gap near $E_F$ in the theory. The solid line in Fig. 2(b) shows the RT fit [22] to the low-*T* data with an effective energy gap of ~9.2 meV, where a satisfactory description of the experiment supports the existence of an energy gap. Second, Kabanov bottleneck model [10,11] is adopted to describe the signal of the gap-like quasiparticles. The corresponding fit to the measured $A_{fast}(T)$ indicates an effective energy gap of 36 meV [line in the inset of Fig. 2(b)]. It should be noted that the adopted models provide a preliminary perception of the order of the gap size even with substantial simplifications of practical circumstance, e.g., momentum and *T* dependences of the gap. The comparable gap sizes revealed from the fast and slow relaxations in FeSe$_{1-x}$ (36 and 9.2 meV), smaller than the similar observations in electron doped iron pnictides (~60 meV) [11], quantitatively support the unified scenario of the gap opening at high temperatures ($T > T_c$). Moreover, we address the onset of the gap opening at 130-140 K where the emergence of gap-like quasiparticles, a small singularity of $\tau_{slow}^{-1}$, and dramatic changes in transport properties (see below) are *simultaneously* observed, as marked by $T_e$ in Fig. 2(b) and its inset [17].

It is worth mentioning the correlation between the orbital modification (eV) and the gap opening (order of 10 meV). Both experimental findings occur at a nearly consistent temperature ~$T_e$, suggesting the same origin of the two features. Before looking into the microscopic causes, it is essential to clarify the evolution of electronic degree of freedom with respect to the elastic ones at $T_s$, where spin fluctuations start to develop [4]. We notice that $T_s$ of FeSe$_{1-x}$ strongly depends on stoichiometry. To exclude inhomogeneity-induced uncertainty in $T_s$, *in situ* observations on the electronic and structural modifications will provide compelling information.

Fig. 3(a) shows the transient optical reflectivity on the sub-nanosecond timescales. Here our concern is an oscillatory feature emerging below 140 K. It becomes more distinct and has a lower oscillation frequency (from 4.2 to 3 GHz, $\hbar\omega$ ~0.01 meV) at lower *T*. Analysis of this feature is performed by fitting with a damped sinusoidal function $A_p\exp(-t/\tau_p)\sin(\omega t + \phi)$. Neither ferromagnetic nor electronic (Rabi) resonances can explain the observed oscillation because of no static magnetic order and large thermal-induced electronic fluctuations. We attribute this signal to the round-trip propagation of photoexcited coherent acoustic phonons, i.e., macroscopic film vibrations, which periodically modulates the thickness and optical property of the FeSe film [23]. Analogous to ultrasound measurements, the vibration frequency reveals the effective longitudinal stiffness $C_{eff} \equiv \rho V_p^2$ [Fig. 3(c)], where $\rho$ and $V_p$ are mass density and the phase velocity of the quasi-longitudinal waves along the [101] direction, respectively [23,24]. Without available low-*T* stiffness of FeSe, we compare the measured $C_{eff}$ at the highest *T* (76 ± 12 GPa at 140 K) with the room-temperature value (77 GPa) [25] and find a reasonable agreement.

The above finding provides valuable clues to both electronic and elastic properties. First, the photoacoustic detection relies on the interference of optical probe beams reflected from the two FeSe interfaces [23], while the room-temperature skin depth of the probe (~26 nm) is shorter than the film thickness by one order of magnitude. The achievement of the acoustic phonon detection, therefore, indicates a significant suppression of the optical absorption below 140 K [Fig. 3(b)], consistent with the distinct spectral weight transfer, as well as orbital modifications, below ~$T_e$ (inset of Fig. 1). Second, the $C_{eff}$ exhibits a *T*-dependent reduction above 90 K and keeps nearly constant at lower *T*. The disclosed phonon softening originates from the lattice instability above the structural phase transition and is arrested at $T_s$, as commonly observed by ultrasonics [26,27]. This is further confirmed by the reference sample with (001) orientation [17]. A *T-independent* $C_{eff}$ (= $C_{33}$) is observed in the temperature range where the phonon signal emerges, agreeing with the picture that the interplane $C_{33}$ is insensitive to the tetragonal-to-orthorhombic distortion. For the study on the (101) film, the coherent phonon dynamics indicates the transition temperature $T_s$ of 90 ± 3 K [Fig. 3(c)], consistent with the x-ray studies on nearly stoichiometric $FeSe_{1-x}$ [13,14] and remarkably below the onset of the electronic modification ($T_e$) by ~50 K.

The *in situ* and stress-free examination serves as an unambiguous basis for unraveling microscopic origin of the high-$T$ gap. Due to the facts of no long-range magnetic order (at all $T$) and monotonic suppression of spin fluctuations above $T_s$ [4], the origin of the high-$T$ gap seems to be some sorts of *short-range* orders in this correlated system where different degrees of freedom are expected to couple together. From the orbital evolution revealed by the spectral weight transfer, we suggest that the gap opening is induced by emergences of short-range orbital and/or charge orders at $T_e$. However, the possibility of a short-range magnetic order cannot be excluded even if it cannot provide direct driving forces for the long-rage lattice distortion. It is noted that the presence of a short-range orbital correlation above $T_s$ in iron pnictides has been predicted by a first principle calculation [28].

From the fact of $T_e > T_s$, we propose a conjecture that the short-range orders evolve into the long-range ones and induce the lattice symmetry breaking at $T_s$. This reminds that the nematic charge order, coupled to the splitting of Fe 3$d$ orbitals, has been observed in iron pnictides below $T_s$ [5]. To examine symmetry of the observed orders, transport properties can provide significant information even if the ordering is short-ranged. In cuprate superconductors, the nematicity was proven to cause a reconstruction of the Fermi surface, leading to dramatic changes in the Hall coefficient $R_H$ and the Seebeck coefficient $S$ [9]. From the inset of Fig. 1, the $R_H$ of $FeSe_{1-x}$ is found to reach the minimum at 150 K and to change the sign at 130 K; meanwhile, a negative maximum of the Seebeck coefficient $S$ is also found in the same high-$T$ region [29]. A refined model, taking into account details of multi-bands of FeSe$_{1-x}$, is necessary for analyzing the $T$-dependent $R_H$ and $S$; but the key point is the striking coincidence of the onset of changes in the transport properties with $T_e$, supporting the scenario that the short-range nematic charge and/or orbital orders induce Fermi surface reconstruction and the gap opening above the structural transition.

The carrier dynamics study provides relevant implications for the high-$T$ anomalies of Fe-based superconductors. First, we notice systematic studies of the structural effect on the cooperation/competition of spin-fluctuation modes and superconductivity [13,30]. This raises the speculation that the structural transition is not simply driven by, but also strongly interacts with the orbital orders in this correlated system. Second, the participation of the high-$T$ gap in electron condensations below $T_c$, argued by a number of experimental works in cuprates and iron pnictides [5,10-12,31], remains puzzling at present. The disclosed strong correlation between the high-$T$ gap and the short-range ordering, therefore, inspires the origin of the precursor order, reminiscent of the recent observations on the nematicity in the pseudogap phase of cuprate compounds [9].

In summary, the *in situ* and stress-free observations on FeSe$_{1-x}$ provide the first evidence for the short-range orders *above* the structural phase transition of Fe-based superconductors. We find a distinct temperature correlation (~140 K) between the transfer of the optical spectral weight, the gap opening, and the onset of changes in the transport properties. These findings reflect the nature of the short-range nematic charge and/or orbital orders that cause Fermi surface reconstruction and the gap opening. This optical spectroscopic study leads to relevant implications for the role of nematicity in the precursor state and the lattice symmetry breaking in high-$T_c$ superconductors.

The authors would like to thank Dr. C. C. Lee, Dr. K.-H. Lin, Prof. T.-K. Lee, and Prof.

# References

*mkwu@phys.sinica.edu.tw

**Figure Captions**

**Fig. 1.** (Color online) Real part of the optical conductivity [$\sigma(\omega) = \sigma_1(\omega) + i\sigma_2(\omega)$]. The arrow depicts the photon energy adopted in the steady-state and the transient reflectivity experiments. The inset shows $T$-dependent optical reflectivity (solid dots) and Hall coefficient $R_H$ (open dots), where the solid line guides to the eye.

**Fig. 2.** (Color online) (a) Picosecond response of transient optical reflectivity $\Delta R/R$ measured at different $T$ (solid lines from top to bottom: 20, 40, 60, 80, 100, 120, 140, 180, 220, 260, 297 K). The data are normalized and vertically shifted for clarity. The dotted lines depict relaxation processes. (b) $T$-dependent $\tau_{slow}^{-1}$ and $A_{fast}$ (dots). The dotted lines are the descriptions of TTM for poor metals in the high-$T$ and low-$T$ limits (HT and LT) with the same $c$-$p$ coupling constant $\lambda$. The solid line is the Rothwarf-Taylor (RT) fit. The line in the inset shows the fitting curve of Kabanov bottleneck model. (c) $T$-dependent fitting variables $A_{slow}$, $A_{step}$, and $\tau_{fast}$.

**Fig. 3.** (Color online) (a) Transient optical reflectivity changes measured at different temperatures (solid lines), which are vertically shifted for clarity. The dashed lines depict the decrease of the oscillation frequency at lower $T$. (b) $T$-dependent oscillation amplitude $A_p$. (c) $T$-dependent effective stiffness. The fitting curve is based on the formula $C_{eff} = C_{eff}^0 (T - \Theta) / (T - T_s)$ [27], where $C_{eff}^0$ is the high-$T$ effective stiffness, and $\Theta$ is a fitting variable.

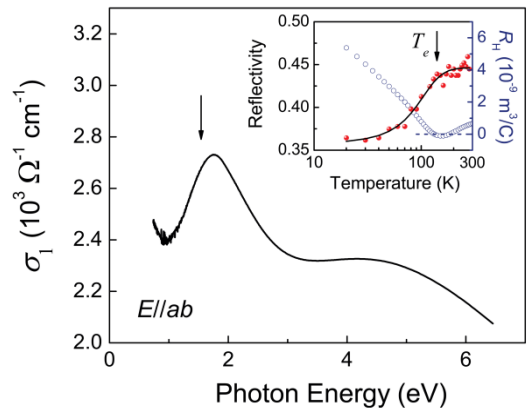

Fig. 1

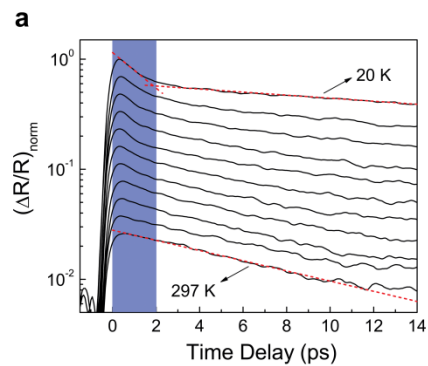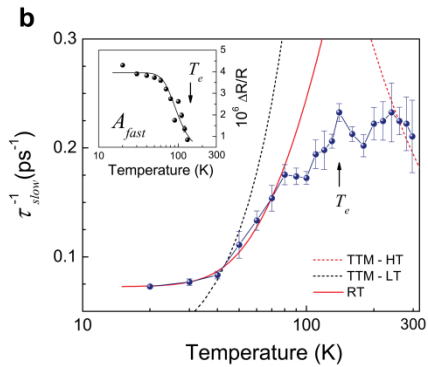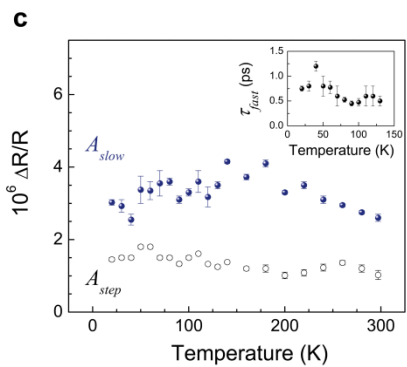

Fig. 2

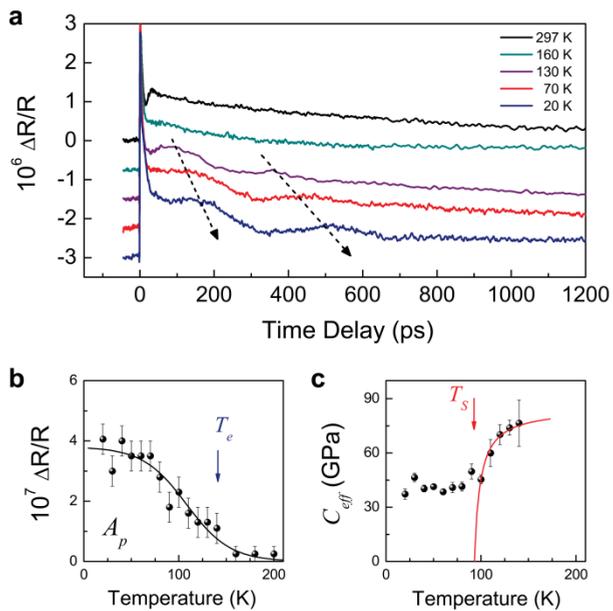

Fig. 3

**EPAPS auxiliary online material for article:**
**Gap opening and orbital modification of superconducting FeSe above the structural distortion**

Y.-C. Wen, K.-J. Wang, H.-H. Chang, J.-Y. Luo, C.-C. Shen, H.-L. Liu, C.-K. Sun, M.-J. Wang, and M.-K. Wu

In this EPAPS auxiliary online document, we include (1) theoretical descriptions of the quasiparticle recombination and carrier-phonon (*c-p*) thermalization in the presence of an energy gap near $E_F$ and carrier-phonon thermalization in metals, (2) transient optical reflectivity measured with the (001) FeSe$_{1-x}$ thin film, and (3) temperature dependence of the amplitude and frequency of coherent acoustic phonon signals observed in the (101) and (001) FeSe thin film.

**(1) Quasiparticle recombination and carrier-phonon thermalization in the presence of an energy gap near $E_F$ and carrier-phonon thermalization in metals**

In metals, the *c-p* thermalization is commonly explained in the framework of the two-temperature model (TTM). This model assumes that the carrier-carrier scattering is much faster than the *c-p* thermalization and considers the energy transfer from quasi-equilibrium carrier to quasi-equilibrium lattice systems. Recently, the model was solved analytically [1]. In the high temperature regime ($T \gg T_D/5$, where $T_D$ is the Debye temperature), the *c-p* relaxation time $\tau$ can be formulated as

$$\tau = \frac{2\pi k_B T}{3\hbar \lambda \langle \omega^2 \rangle}, \tag{1}$$

where $k_B$ is the Boltzmann constant, and $\lambda \langle \omega^2 \rangle$ is the second moment of the Eiashberg function. We use Eq. (1) to fit the measured *c-p* relaxation rate $\tau_{slow}^{-1}$ in the high-temperature regime with $\langle \omega^2 \rangle = (27 \text{ meV})^2$ [2]. For the (001) thin film, where the probe polarization is parallel to the *ab* plane, we obtained a *c-p* coupling constant $\lambda = 0.13$, reasonably agreeing with the calculation value ($\lambda = 0.17$, equivalent to a $\tau_{slow}^{-1} = 3.5$ ps$^{-1}$) [2]. (The experimental data is shown below). For the (101) thin film, we obtained a $\lambda \sim 0.008$. The corresponding fitting curve and the interpretation of the anisotropic carrier dynamics are given in Fig. 2(b) and content of the manuscript.

In the low temperature regime, the *c-p* thermalization exhibits a very different trend. We consider the case of a poor metal for the studied FeSe thin film [1]. The corresponding *c-p* relaxation time can be formulated as

$$\tau = \frac{2\hbar^2 T_D}{\pi^3 \lambda (k_B T)^2}. \tag{2}$$

This model allows for estimating the *c-p* relaxation time at low temperatures. The corresponding model prediction with $\lambda = 0.008$ and $T_D = 289$ K [3] is also shown in Fig. 2(b) of the manuscript. A clear discrepancy between this theoretical curve and the experiments is found and cannot be alleviated by considering the case of clean metals ($\tau \propto T^{-3}$).

To understand the effect of an energy gap on the *c-p* thermalization, Kabanov *et. al.* analytically solved the Rothwarf-Taylor (RT) theory [4] and provided the expression of the modified *c-p* relaxation time [5]:

$$\tau = \frac{\tau_0}{\delta (\varepsilon n_T + 1)^{-1} + 2n_T}, \quad (3)$$

where $n_T = \sqrt{T} \exp(-\Delta/T)$ is a thermal quasiparticle density, and $\tau_0$, $\delta$, $\varepsilon$, and $\Delta$ are temperature-independent fitting variables. With the substantial simplifications of the practical circumstance, this phenomenological model enables a rough estimation of the effective energy gap $\Delta$. We adopt Eq. (3) to describe the measured $\tau_{slow}^{-1}$ at low temperatures where the gap participates in the thermalization, as shown by the solid line in Fig. 2(b). The yield fitting variables are $\tau_0 \sim 33$ ps, $\delta = 0.4$, $\varepsilon = 0.8$, and $\Delta = 107 \pm 17$ K.

The gap effect is also revealed by the emergence of the gap-like quasiparticles, i.e., the nonequilibrium carriers accumulating in the quasiparticle state above the gap and waiting for recombination via scatterings with high-energy phonons. Kabanov *et. al.* proposed a model to deal with this bottleneck effect [6], enabling quantitative analyses of the effective gap size $\Delta$. Here we assume a *T*-independent $\Delta$ for a preliminary estimation of the effective gap size. Under the weak perturbative condition, the signal amplitude of gap-like quasiparticles can be formulated as $A_{fast} \propto [1 + B \exp(-\Delta/kT)]^{-1}$, where $B = 2\nu/N(0)\hbar\Omega_c$. $\nu$ is the number of phonon modes interacting with the quasiparticles, $N(0)$ is the electronic density of states near $E_F$, and $\Omega_c$ is a typical phonon cutoff frequency. With $N(0) = 1.29$ state eV$^{-1}$ Fe$^{-1}$ and $\Omega_c = 350$ cm$^{-1}$ for FeSe [2,7], we can fit the measured $A_{fast}$, yielding the fitting variables $\nu = 1.6\sim2$ and $\Delta = 36 \pm 4$ meV.

It should be noted that the adopted models under substantial simplifications provide a preliminary perception of the order of the gap size. These theories are, however, unable to determine exact *T* dependence of the gap size and the gap anisotropy. In this presented work, we address the onset of the gap opening at 130-140 K based on the facts of several coincident experimental features, including the emergence of gap-like quasiparticles, a small singularity of *c-p* thermalization rate, and the onset of changes in the transport properties.

**(2)  Transient optical reflectivity measured with the (001) FeSe1-x thin film**

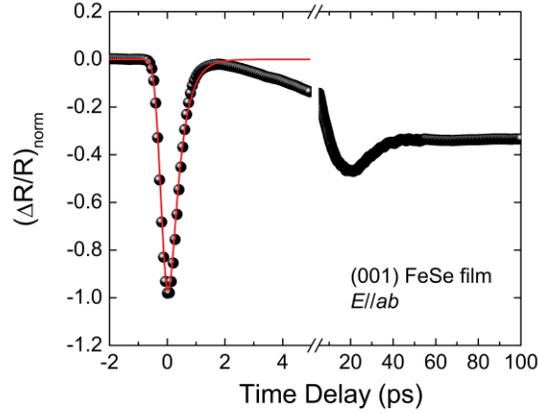

The solid dots show the ΔR/R measured with the (001) FeSe thin film at room temperature, with the probe polarization parallel to the *ab* plane. The red curve is a fitting curve with a time constant of 370 fs. The dynamics on the longer timescales reveals the heat diffusion and the coherent acoustic phonon dynamics.

**(3)  Temperature dependence of the amplitude and the frequency of coherent acoustic phonon signals observed in the (101) and (001) FeSe thin film**

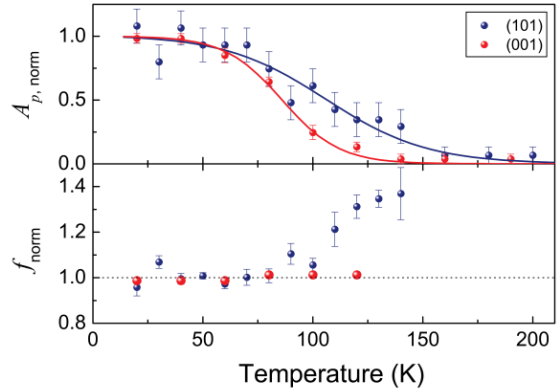

Amplitude $A_p$ and frequency $f$ of the coherent acoustic phonon oscillations observed in the (101) and (001) FeSe thin films. Both $A_p$ and $f$ are normalized with respect to the low-$T$ values.

The (101) data are the same as these shown in Fig. 3 of the manuscript. In case of the (001) thin film, coherent phonon signal emerges below 120-140 K, and its amplitude increases with decreasing $T$ until ~60 K. In this temperature region, no identifiable change in the oscillation frequency, as well as the corresponding $C_{33}$ (= $C_{eff}$), can be found. This result agrees with the interpretation that interplane $C_{33}$ does not exhibit remarkable softening as the crystal undergoes tetragonal-to-orthorhombic structural transition.

It was noted that the (001) experiment used for comparing anisotropic lattice softening was not carried out under the weak perturbative condition. The stronger pump fluence made the error bars of the (001) data smaller, while the steady-state sample heating caused a shift of the (001) data in comparison with the (101) ones by ~20 K. This deficiency does not affect the conclusion taken from this experiment and all statements in the paper. A more delicate investigation on $T$-dependent $C_{33}$ using a pulse-and-echo scheme can be found in our earlier publication [3]. In that picosecond ultrasonic work, the surface oxidation restricted the optical penetration within ~30 nm near the surface, and thus the data showed clear coherent acoustic phonon echoes.